\documentclass[aps,preprint]{revtex4}

\usepackage{graphicx}

\newcommand{\be}{\begin{equation}}
\newcommand{\ee}{\end{equation}}

\begin{document}

\title{\bf Computation of Neutron Star Structure\\
Using Modern Equation of State}
\author{G.H. Bordbar\footnote{Corresponding author}
\footnote{E-Mail: bordbar@physics.susc.ac.ir} and M. Hayati}

\affiliation{Department of Physics, Shiraz University, Shiraz
71454, Iran\footnote{Permanent address}\\
and\\
Research Institute for Astronomy and Astrophysics of Maragha,\\
P.O. Box 55134-441, Maragha, Iran }
%************************************************************************
\begin{abstract}
Using the modern equations of state derived from microscopic
calculations, we have calculated the neutron star structure. For
the neutron star, we have obtained a minimum mass about $0.1\ {\rm
M_{\odot}}$ which is nearly independent of the equation of state,
and a maximum mass between $1.47\ {\rm M_{\odot}}$ and $1.98\ {\rm
M_{\odot}}$ which is strongly dependent on the equation of state.
It is shown that among the equations of state of neutron star
matter which we have used, the stiffest one leads to higher
maximum mass and radius and lower central density. It is seen that
the given maximum mass for the Reid-93 equation of state shows a
good consistency with the accurate observations of radio pulsars.
We have indicated that the thickness of neutron star crust is very
small compared to the predicted neutron star radius.
\end{abstract}
%\noindent{keywords: Equation of state, structure, neutron star}
%************************************************************************
\maketitle
\section{Introduction}
\label{intro} For identifying of an astrophysical object as a
black hole, it is required to know the maximum gravitational mass
of a neutron star for stability against collapse into a black
hole. In the other word, it is expected that below a certain
maximum mass, degeneracy pressure due to the nucleons is
sufficient to prevent an object from becoming a black hole
\cite{Shapiro83}. Therefore, determining the maximum gravitational
mass of neutron star is of special importance in astrophysics.

Observationally, the X-ray pulsars and X-ray bursters can offer
direct measurement of the neutron star mass, but their accuracy is
rather poor and due to large errors, measuring masses of them are
not very useful. Fortunately, the masses of neutron stars have
been determined with high accuracy using the binary radio pulsars
 \cite{Weisberg84,Liang86,Fujimoto86,Taylor89,Heap92,Brown96,
 Miller98,Thorsett99,Orosz99,Lattimer01,Heiselberg02,Jonker03,
 Haensel03,Quaintrell03}.

Theoretically, in calculating a maximum mass of neutron star which
should be in agreement with the precise observations, the equation
of state of neutron star matter plays a crucial role. Theoretical
investigations of the neutron star matter indicates that a
considerable uncertainties exist in the behavior of the equation
of state, especially at high densities. This is due to using the
old non phase-shift equivalent nucleon-nucleon potentials
\cite{Engvik97}. These uncertainties result a significant
uncertainty in the maximum mass of neutron star. Therefore having
a modern equation of state derived from an accurate many-body
calculation using modern nucleon-nucleon potentials is of
particular importance in the mass determination for the neutron
star \cite{Lattimer00,Lattimer01,Haensel03}.

Recently, we have obtained the equation of state of neutron star
matter using the microscopic constrained variational calculations
based on the cluster expansion of the energy functional
\cite{Bordbar02,Bordbar04}. In these calculations, we have
employed the modern two-nucleon potentials such as the new Argonne
${\rm AV}_{18}$ \cite{Wiringa95} and charged dependent Reid-93
\cite{Stoks94}. This is a fully self-consistent technique which
does not bring any free parameter into the calculations and  its
results show a good convergence. In this method, for the
asymmetric matter calculations, a microscopic computation of
asymmetry energy is done and therefore its results are more
accurate with respect to the other methods which use the
semi-empirical parabolic approximation. In fact, using the modern
nucleon-nucleon potentials which are explicitly depend on the
isospin projection ($T_z$), requires a microscopic calculation
\cite{Bordbar97,Bordbar98,Bordbar03}. In this article, we
investigate some physical properties of neutron star structure
using modern equations of state of neutron star matter
\cite{Bordbar02,Bordbar04}.
%%%%%%%%%%%%%%%%%%%%%%%%%%%%%%%%%%%%%%%%%%%%%%%%%%%%%%%%%%%%%%%%%%%%%%%%%%
\section{Equation of State and Neutron Star Structure}
\begin{figure}

 \includegraphics[height=15cm]{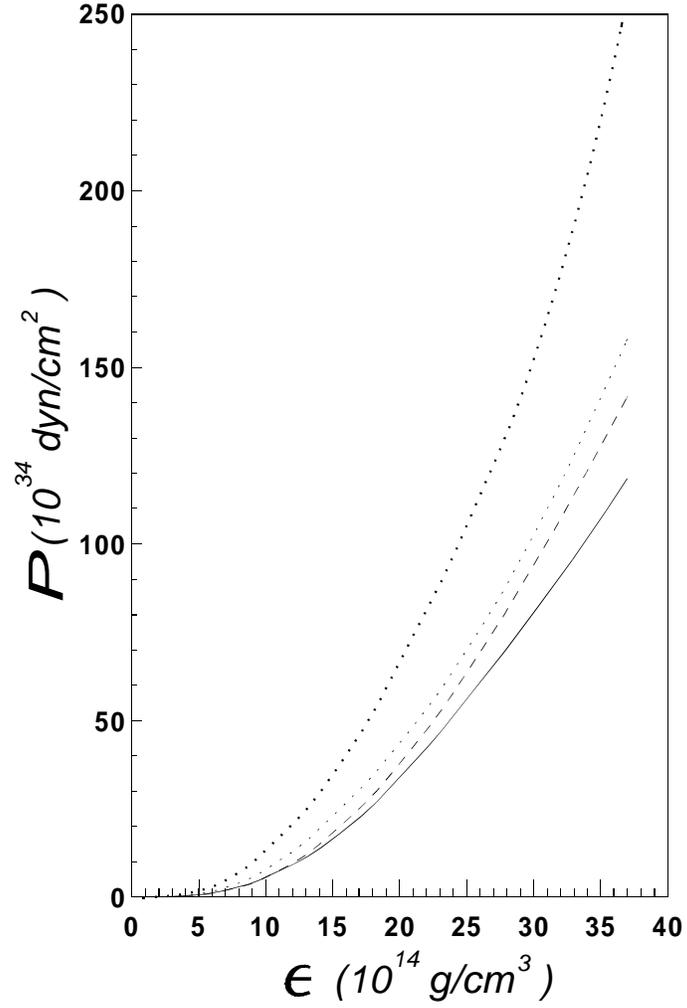}

 \caption{
Pressure ($P$) versus mass density ($\epsilon$) for the neutron
star matter with the Reid-93 (full curve), ${\rm AV}_{14}$ (dashed
curve), ${\rm AV}_{18}$ (dotted curve) and ${\rm UV_{14}+TNI}$
(heavy dotted curve) potentials.
  }

  \label{fig1}
  \end{figure}

\begin{figure}

\includegraphics[height=15cm]{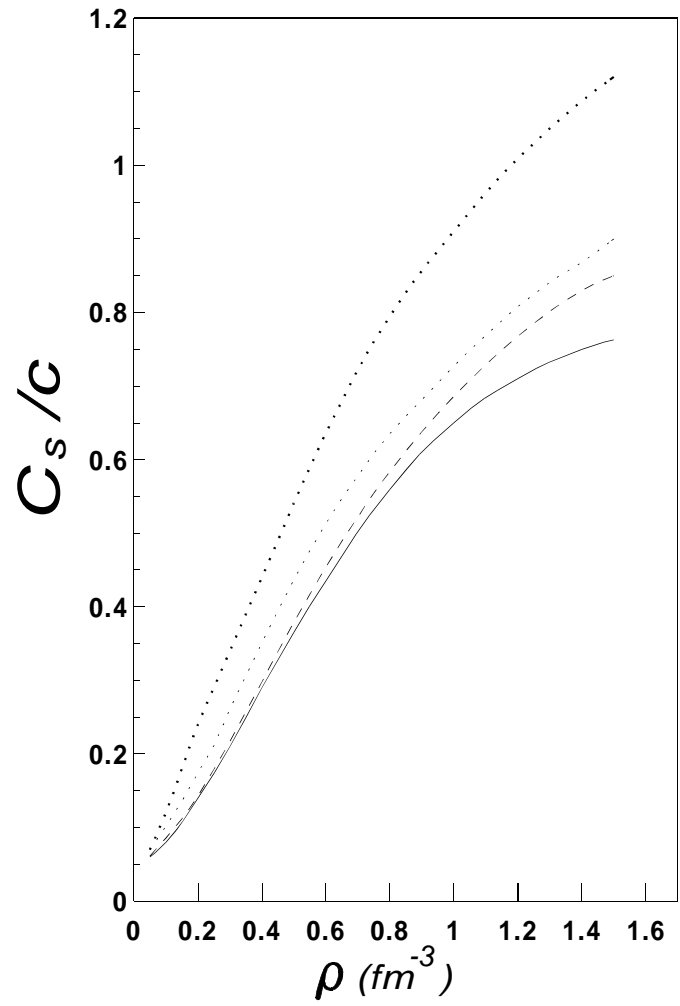}

\caption{ As Figure~\ref{fig1}, but for the sound speed ($C_{\rm
s}/c$) versus number density ($\rho$).} \label{fig2}
\end{figure}

As it is mentioned in the previous section, the equation of state
of neutron star matter has a key role in determining the neutron
star structure, especially its maximum mass. In this work, we
study the structure of neutron star using the modern microscopic
equations of state of neutron star matter employing the ${\rm
AV}_{18}$ and Reid-93 modern two-nucleon potentials as well as the
${\rm AV}_{14}$ potential \cite{Wiringa84}. In our calculations,
we also consider the effect of three-body force using the ${\rm
UV_{14}+TNI}$ potential in which the effect of three-nucleon
interaction is included by adding two density dependent terms into
the two-body potential \cite{Lagaris81}. The procedure of our
calculations is discussed in references
\cite{Bordbar02,Bordbar04}. Our results for the equation of state
of neutron star matter are given in Figure~\ref{fig1}. It is seen
that by inclusion of three-nucleon interaction, we find a stiffer
equation of state. From Figure~\ref{fig2}, it is seen that for the
Reid-93, ${\rm AV}_{14}$ and ${\rm AV}_{18}$ equations of state,
the speed of sound ($C_{\rm s}$) is always less than the speed of
light, even at high densities. In the case of ${\rm UV_{14}+TNI}$
equation of state, above $\rho = 1.28\ {\rm fm}^{-3}$ ($\epsilon =
29.61\times 10^{14}\ {\rm g/cm^3}$), the sound speed exceeds the
speed of light. However, as we will see, for this equation of
state, the neutron star mass reaches a limiting value (maximum
mass) below this density.

Using the equation of state of the neutron star matter, we can
calculate the neutron star mass and radius as a function of
central mass density, $\epsilon_{\rm c}$, by numerically
integrating the general relativistic equation of hydrostatic
equilibrium, Tolman-Oppenheimer-Volkoff (TOV) equation
\cite{Shapiro83},
\begin{equation}
\frac{dP}{dr}=-\frac{G}{r^2}[\epsilon (r) +P(r)/c^2]
\frac{m(r)+4\pi r^3P(r)/c^2}{1-\frac{2Gm(r)}{rc^2}},
\end{equation}
where
\begin{equation}
\epsilon =\rho [E(\rho)+mc^2]
\end{equation}
is the mass density, $G$ is the gravitational constant, and
\begin{equation}
m(r) =\int_0^r 4\pi r'^2\epsilon (r')dr'
\end{equation}
has the interpretation of the mass inside radius $r$. By selecting
a central mass density $\epsilon_{\rm c}$, under the boundary
conditions $P(0)=P_{\rm c}$,  $m(0)=0$, we integrate the TOV
equation outwards to a radius $r=R$, at which $P$ vanishes. This
yields the neutron star radius $R$ and mass $M=m(R)$.

For densities greater than $0.05\ {\rm fm}^{-3}$, we use our
equations of state presented in Figure~\ref{fig1}. However, at the
lower densities, we use the equation of state calculated by Baym
et al. \cite{Baym71}, since the details of the equation of state
at low densities do not affect our results.

\begin{figure}

\includegraphics[height=15cm]{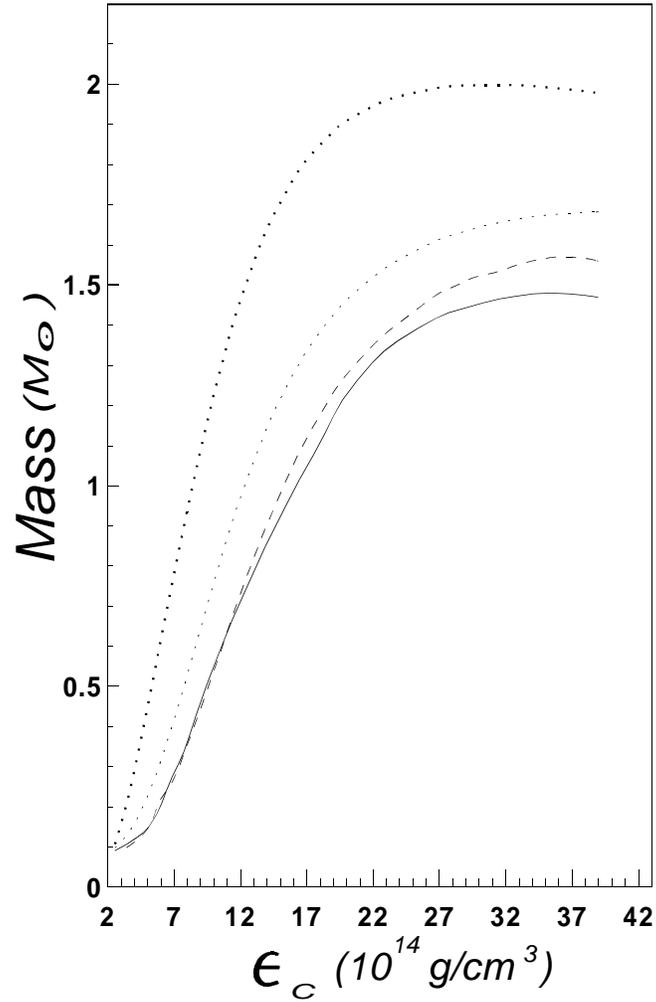}

\caption{ As Figure~\ref{fig1}, but for the gravitational mass of
neutron star (in units of the solar mass ${\rm M_{\odot}}$) versus
central mass density ($\epsilon_{\rm c}$).}
 \label{fig3}
\end{figure}

The neutron star gravitational mass (in solar mass ${\rm
M_{\odot}}$ units) as a function of central mass density,
$\epsilon_{\rm c}$, calculated with the ${\rm AV}_{18}$, Reid-93,
${\rm AV}_{14}$ and ${\rm UV_{14}+TNI}$ equations of state is
shown in Figure~\ref{fig3}. Our results show that at low
densities, the calculated neutron star masses exhibit a minimum
($\simeq 0.1\ {\rm M_{\odot}}$) which depends weakly on the
equation of state of neutron star matter. It can be seen that at
high densities, the increasing of mass becomes very slow and
finally it approaches a limiting value which strongly depends on
the equation of state. This limiting value of mass is the maximum
gravitational mass of neutron star and its corresponding central
density is the highest possible value for the neutron star central
density. A star with the higher central density would be unstable
against the gravitational collapse to a black hole. A comparison
between Figure~\ref{fig1} and Figure~\ref{fig3} indicates that
there is a one-to-one correspondence between the behavior of
equation of state of neutron star matter, especially at high
densities, and the resulting mass of neutron star. The stiffest
equation of state leads to the higher mass and lower central
densities. It should be noted that, as mentioned above, the
limiting value of neutron star mass with the ${\rm UV_{14}+TNI}$
equation of state is reached below the density in which the speed
of sound exceeds the speed of light.

\begin{figure}

\includegraphics[height=15cm]{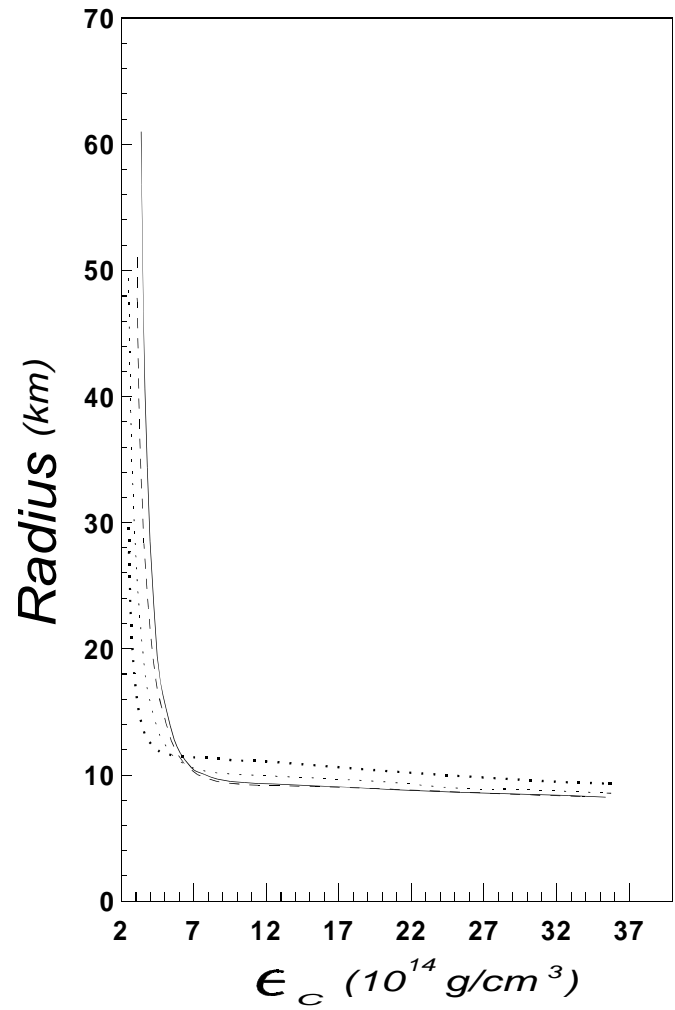}

\caption{ As Figure~\ref{fig1}, but for the radius versus central
mass density ($\epsilon_{\rm c}$).} \label{fig4}
\end{figure}

In Figure~\ref{fig4}, We have presented the radius of neutron star
with the ${\rm AV}_{18}$, Reid-93, ${\rm AV}_{14}$ and ${\rm
UV_{14}+TNI}$ potentials as a function of central mass density. It
is seen that as the central density increases, the radius
decreases very rapidly and then for $\epsilon_{\rm c}\geq 7\times
10^{14}\ {\rm g/cm^3}$, it reaches a nearly constant value. This
shows that the radius of neutron star is nearly independent of the
central density. From Figure~\ref{fig4}, we also observe that the
stiffest equation of state leads to a relatively larger radius for
the neutron star.

\begin{figure}

\includegraphics[height=15cm]{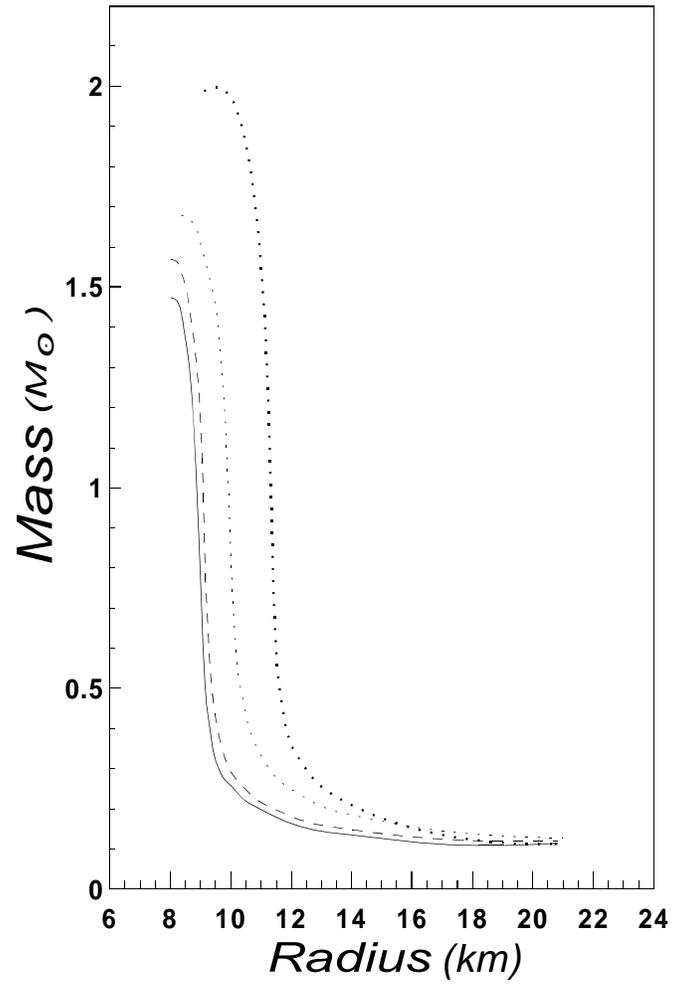}

\caption{ As Figure~\ref{fig1}, but for the mass-radius relation.}
 \label{fig5}
\end{figure}

The gravitational mass versus radius (M-R curve) for the ${\rm
AV}_{18}$, Reid-93, ${\rm AV}_{14}$ and ${\rm UV_{14}+TNI}$
potentials is shown in Figure~\ref{fig5}. Our results indicate
that for the neutron star, there is a minimum gravitational mass
about $0.1\ {\rm M_{\odot}}$ which is nearly identical for
different equations of state, and a maximum mass which is
different for these equations of state. It is well known that in
this mass region, the equilibrium configuration of neutron stars
can exist.

A summary of our results for the properties of maximum mass
configuration of neutron star predicted for different equations of
state are given in Table~\ref{tab1}. From Table~\ref{tab1}, it can
be seen that the inclusion of the three-nucleon interaction
considerably affects the calculated properties of neutron star
maximum mass configuration. Here, we explicitly see that the
stiffest equation of state used in our calculations gives the
lower central density and higher maximum mass and radius for the
neutron star. As it is seen in Table~\ref{tab1}, our results give
the maximum mass of neutron star between $1.47\ {\rm M_{\odot}}$
and $1.98\ {\rm M_{\odot}}$. This agrees with the measured range
of neutron star masses and the results of other theoretical
investigations \cite{Haensel03,Lattimer01,Wiringa88,Baldo97}.
Furthermore, our predicted maximum mass of neutron star with the
Reid-93 potential has a good consistency with the mass determined
from the accurate observations of radio pulsars \cite{Weisberg84,
Taylor89,Thorsett99}. This is a good confirmation for our equation
of state of dense matter with the modern nucleon-nucleon
potential, Reid-93. If we use the equation of state of
noninteracting beta-stable matter, a maximum mass $0.7\ {\rm
M_{\odot}}$ is obtained. Comparing this value with those presented
in Table~\ref{tab1} implies that at high densities, the
nucleon-nucleon interaction is sufficiently repulsive to produce a
considerable shift in the maximum mass of neutron star with
respect to the noninteracting case.

For a neutron star, the binding energy is defined as the energy
needed to transform it into a dispersed configuration. Formally,
the binding energy can be obtained from the difference between amu
mass $M_{\rm A}$ and gravitational mass, where $M_{\rm A}$ is
given by \cite{Wiringa88}
\begin{equation}
M_{\rm A}= m_{\rm A}\int_0^R\frac{4\pi r^2\epsilon
(r)}{[1-2m(r)G/rc^2]^{1/2}}dr\cdot
\end{equation}
Here, we take $m_{\rm A}$ as one amu ($1.66\times 10^{-24}\ {\rm
g}$). For all values of gravitational masses (greater than the
predicted minimum mass), our equations of state give positive
binding energies for neutron stars. This indicates that the
neutron stars are always stable against the dispersed
configuration.

\begin{figure}

\includegraphics[height=15cm]{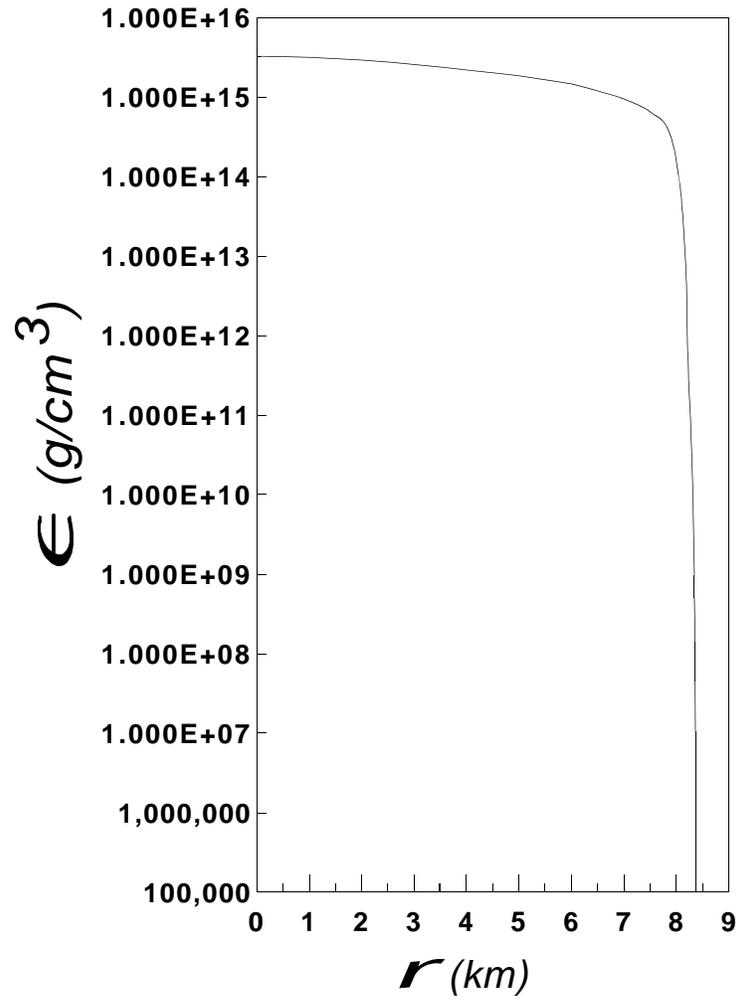}

\caption{ Mass density as a function of radial coordinate for a
$1.47\ {\rm M_{\odot}}$ neutron star with the Reid-93 potential.}
\label{fig6}
\end{figure}

We have calculated the mass density versus radial coordinate for a
neutron star with the mass equal to $1.47\ {\rm M_{\odot}}$ for
the Reid-93 equation of state. This is presented in
Figure~\ref{fig6} which shows that up to $r\simeq 7.8\ {\rm km}$
(within the core), the mass density is nearly uniform, and then it
drops very rapidly (close to the crust). Our results indicate that
the crust thickness is about $0.43\ {\rm km}$ which is only $5.2
\%$ of our predicted neutron star radius. This implies that the
core contains the most fraction of neutron star matter.

By solving the TOV equation, we have also calculated the
properties of neutron star with the gravitational mass equal to
$1.4\ {\rm M_{\odot}}$. Our results for different equations of
state are presented in Table~\ref{tab2}. Comparing
Tables~\ref{tab1} and \ref{tab2} indicates that for a $1.4\ {\rm
M_{\odot}}$ neutron star, our used equations of state give higher
radius and lower central density with respect to those predicted
for maximum mass configuration. The differences between these two
cases become more significant for the stiffest equation of state.

\section{Summary and Conclusion}
In recent years, a wide range of information on neutron stars are
available from X-ray pulsars, X-ray bursters and radio pulsars.
Therefore, the study of neutron stars structure on the basis of
the equation of state of dense matter is of special interests in
astrophysics. Theoretical investigation of neutron star structure
is also important, since the observational results leads to the
constraints on the equation of state of dense matter.

In present paper, we have computed some properties of neutron star
structure using the modern equations of state of neutron star
matter obtained by microscopic constrained variational
calculations based on the modern potentials. A minimum value for
the gravitational mass of neutron star is predicted near $0.1\
{\rm M_{\odot}}$ which is nearly identical for different equations
of state. It is seen that a higher maximum mass is obtained, when
we consider the effect of three-nucleon interaction in our
calculations. We have shown that the radius of neutron star
becomes nearly constant for all central densities approximately
greater than $7\times 10^{14}\ {\rm g/cm^3}$. The properties of
maximum mass configuration of neutron star are predicted. These
properties change considerably by considering the three-nucleon
interaction. Our results show that the stiffest equation of state
gives the higher maximum mass and radius and lower central density
for the neutron star. The obtained maximum mass of neutron star is
between $1.47\ {\rm M_{\odot}}$ and $1.98\ {\rm M_{\odot}}$ which
agrees with the masses determined from observations. In addition,
the predicted value $1.47\ {\rm M_{\odot}}$ by our used Reid-93
equation of state is in a good agreement with the accurately
determined mass of radio pulsars. We have seen that in our
calculated mass region, the binding energy of neutron star is
positive and therefore it is stable with respect to the dispersed
configuration. We have computed the mass density profile for a
neutron star with the mass equal to $1.47\ {\rm M_{\odot}}$ for
the Reid-93 equation of state. It is shown that the neutron star
crust has a thickness about $0.43\ {\rm km}$ which is very small
with respect to the obtained neutron star radius. We have also
computed the properties of a $1.4\ {\rm M_{\odot}}$ neutron star
with the different equations of state. A higher radius and lower
central density are predicted for a $1.4\ {\rm M_{\odot}}$ neutron
star compared to the values calculated for the maximum mass
configuration. Finally, a good agreement between our results for
the neutron star structure and those of other theoretical
calculations is observed.
%%%%%%%%%%%%%%%%%%%%%%%%%%%%%%%%%%%%%%%%%%%%%%%%%%%%%%%%%%%%%%%%%%%%%%%%%%
\acknowledgements {This work has been supported by Research
Institute for Astronomy and Astrophysics of Maragha, and Shiraz
University Research Council.}
%%%%%%%%%%%%%%%%%%%%%%%%%%%%%%%%%%%%%%%%%%%%%%%%%%%%%%%%%%%%%%%%%%%%%%%%%%%

%%%%%%%%%%%%%%%%%%%%%%%%%%%%%%%%%%%%%%%%%%%%%%%%%%%%%%%%%%%%%%%%%%%
\newpage
\begin{table}
\begin{center}
\caption{ Maximum gravitational mass ($M_{\rm max}$),
corresponding radius ($R$), central mass density ($\epsilon_{\rm
c}$), and central number density ($\rho_{\rm c}$) obtained with
the ${\rm AV}_{18}$, Reid-93, ${\rm AV}_{14}$ and ${\rm
UV_{14}+TNI}$ potentials.}
\begin{tabular}{ccccc}
\hline Potential&$M_{\rm max}\ ({\rm M_{\odot}})$&$R\ {\rm (km)}$&
$\epsilon_{\rm c}\ (10^{14}\ {\rm g/cm^3})$&
$\rho_{\rm c}\ ({\rm fm^{-3}})$\\
\hline
Reid-93&1.47&8.23&35.39&1.7\\
${\rm AV}_{14}$&1.56&8.28&33.93&1.6\\
${\rm AV}_{18}$&1.65&8.79&31.16&1.45\\
${\rm UV_{14}+TNI}$&1.98&9.81&27.17&1.2\\
\hline
\end{tabular}
\end{center}
\label{tab1}
\end{table}
%***********************************************************************
\begin{table}
\begin{center}
\caption{ Properties of neutron star with the gravitational mass
equal to $1.4\ {\rm M_{\odot}}$ for different equations of state.}
\begin{tabular}{cccc}
\hline Potential&$R\ {\rm (km)}$&$\epsilon_{\rm c}\ (10^{14}\ {\rm
g/cm^3})$&
$\rho_{\rm c}\ ({\rm fm^{-3}})$\\
\hline
Reid-93&8.51&27.71&1.4\\
${\rm AV}_{14}$&8.81&22.83&1.2\\
${\rm AV}_{18}$&9.58&18.25&0.95\\
${\rm UV_{14}+TNI}$&11.16&11.32&0.6\\
\hline
\end{tabular}
\label{tab2}
\end{center}
\end{table}
%%%%%%%%%%%%%%%%%%%%%%%%%%%%%%%%%%%%%%%%%%%%%%%%%%%%%%%%%%%%%%%%%%%
\newpage
%%%%%%%%%%%%%%%%%%%%%%%%%%%%%%%%%%%%%%%%%%%%%%%%%%%%%%%%%%%%%%%%%%%%%%%
%%%%%%%%%%%%%%%%%%%%%%%%%%%%%%%%%%%%%%%%%%%%%%%%%%%%%%%%%%%%%%%%%%%%%%%
%%%%%%%%%%%%%%%%%%%%%%%%%%%%%%%%%%%%%%%%%%%%%%%%%%%%%%%%%%%%%%%%%%%
%%%%%%%%%%%%%%%%%%%%%%%%%%%%%%%%%%%%%%%%%%%%%%%%%%%%%%%%%%%%%%%%%%%
%********************************************************************
%********************************************************************

%********************************************************************
\end{document}